\documentclass[
aps,%
12pt,%
final,%
notitlepage,%
oneside,%
onecolumn,%
nobibnotes,%
nofootinbib,
superscriptaddress,%
noshowpacs,%
centertags]%
{revtex4}

\begin{document}
\selectlanguage{english} 
\title{Three-Dimensional Two-Pion Emission Source at SPS: \\ Extraction of Source Breakup Time and Emission Duration}
\author{\firstname{P.} \surname{Chung}}
\email[]{pchung@mail.chem.sunysb.edu}
\affiliation{Dept of Chemistry, SUNY Stony Brook, Stony Brook, NY 11794, USA}
\author{\firstname{P.} \surname{Danielewicz}}
\email[]{danielewicz@nscl.msu.edu}
\affiliation{National Superconducting Cyclotron Laboratory, MSU, East Lansing, MI, USA,\\
Dept of Physics \& Astronomy, Michigan State University, East Lansing, MI, USA.}
\author{the NA49 Collaboration}
\begin{abstract}
A model-independent, three-dimensional source function for pion pairs has been
extracted from Pb+Pb collisions at $\sqrt s_{NN}=17.3$~AGeV. The extracted source exhibits long-range non-Gaussian tails in the directions of the pion-pair net transverse-momentum and of the beam. Comparison with the Therminator model allows for an extraction of the pion source proper breakup time and of emission duration in the collisions.

\end{abstract}
\maketitle
\section{Introduction}

      A deconfined phase of nuclear matter is expected to be formed at the
 high energy densities created in relativistic heavy ion collisions \cite{qgp06}. It is
 expected that the signatures of such a phase
 are reflected in the space-time extent and shape of particle emission
 source-functions.

     Recently, one-dimensional (1D) source imaging techniques \cite{brown97,brown98,brown01} have revealed a
 non-trivial long-range structure in the two-pion emission source at RHIC
\cite{chung05,chung06}. The origins of this structure are still investigated.
The nature of such structures and the potentially useful information they can provide could be revealed, on one hand, by examining the possible presence of such a structure in heavy-ion collisions at intermediate SPS energies and, on the other, by examining the dependence of a structure on the direction within the source. The NA49 Collaboration has investigated Pb+Pb collisions over a wide range of bombarding energies at the CERN SPS during the last decade \cite{alt05}, accumulating pion data.  The rich 
data set provides a unique opportunity to search for non-trivial structures at the SPS, investigate their dependence on energy and direction and extract any useful information those structures may provide.

In this paper, the first three-dimensional (3D) emission source image for low $p_T$ pion
pairs produced in central Pb+Pb collisions at $\sqrt s_{NN} = 17.3$~GeV is presented. The result is compared with calculations from the Therminator model in order to extract the source proper breakup time and emission duration.

\section{Setup and Analysis}

\subsection{Experimental Setup}

The data presented here were taken by the NA49 Collaboration during the years 1996-2002. The incident beams of 158 AGeV were provided by the CERN SPS accelerator. The NA49 Large Acceptance Hadron Detector \cite{afa99}
achieves a large-acceptance precision-tracking ($\delta p/p^2 \approx (0.3-7).10^{-4}$ (GeV/c$)^{-1}$) and particle identification using time projection chambers
(TPC's).
Charged particles are detected by the tracks left in the TPC's and identified by the energy deposited in the TPC gas. Mid-rapidity particle identification is further enhanced by a time-of-flight wall (resolution of 60~ps). Event centrality is
determined by a~forward calorimeter which measures the energy of spectator matter.

\subsection{Three-Dimensional Correlation Function}

The 3D correlation function, C($\mathbf{q}$), was calculated as the ratio of pair to uncorrelated reference distributions in relative momentum $\mathbf{q}$ for  $\pi^+\pi^+$ and $\pi^-\pi^-$ pairs. Here, $\mathbf{q}=\frac{(\mathbf{p_1}-\mathbf{p_2})}{2}$ 
is half of the momentum difference between the two particles in the pair Center-of-Mass System (PCMS) frame. The pair
distribution was obtained using pairs of particles from the same event and the uncorrelated distribution
was obtained by pairing particles from different events. The Lorentz transformation of $\mathbf{q}$ from the laboratory frame to the PCMS was done by a Lorentz transformation to the pair Locally Co-Moving System (LCMS) frame along the beam
direction followed by a Lorentz transformation to the PCMS along the direction of the transverse momentum of the pair.

Track merging and splitting effects were suppressed by appropriate cuts on both the pair and uncorrelated distributions. The pair cuts require the two particles in each pair to be separated by at least 2.2 cm over 50 pad rows in the vertex TPC's. A 20$\%$ increase in this minimum separation results in the correlation data points fluctuating within the statistical errors. Hence, the systematic uncertainty associated with the pair cuts is deemed smaller than the statistical uncertainty.

The effects of track momentum resolution were assessed by jittering the momentum of the tracks in the data by the maximum momentum resolution, $\delta p/p^2 \approx 7.10^{-4}$ (GeV/c$)^{-1}$, and by re-computing the 3D correlation function. The resulting 
correlation function incorporates then twice the effect of the momentum resolution and was found to be very close to the raw un-smeared correlation function. This is not surprising, considering that the mean momentum of the tracks used in this analysis is
 1.4 GeV/c, resulting in a momentum resolution of $\delta p/p \approx 0.1\%$. The above procedure gives a very small smear to the correlation peak in the region $q<10$~MeV.  In practice, this tends to reduce the imaged source intensity at large separation. As a consequence, the results for the source function presented here may slightly
underestimate the actual source function at large $r$.

\subsection{Angular Moments}

In the Cartesian surface-spherical harmonic decomposition technique \cite{daniel05,chung05}, the 3D correlation function is expressed as
\begin{equation}
C(\mathbf{q}) - 1 = R(\mathbf{q}) = \sum_l \sum_{\alpha_1 \ldots \alpha_l}
   R^l_{\alpha_1 \ldots \alpha_l}(q) \,A^l_{\alpha_1 \ldots \alpha_l} (\Omega_\mathbf{q})
 \label{eqn1}
\end{equation}
where $l=0,1,2,\ldots$, $\alpha_i=x, y \mbox{ or } z$, $A^l_{\alpha_1 \ldots \alpha_l}(\Omega_\mathbf{q})$
are Cartesian harmonic basis elements ($\Omega_\mathbf{q}$ is solid angle in $\mathbf{q}$ space) and $R^l_{\alpha_1 \ldots \alpha_l}(q)$ are Cartesian correlation moments given by
\begin{equation}
 R^l_{\alpha_1 \ldots \alpha_l}(q) = \frac{(2l+1)!!}{l!}
 \int \frac{d \Omega_\mathbf{q}}{4\pi} A^l_{\alpha_1 \ldots \alpha_l} (\Omega_\mathbf{q}) \, R(\mathbf{q})
 \label{eqn2}
\end{equation}
where $q$ is the modulus of the 4-vector $\mathbf q$.
The coordinate axes are oriented so that $z$ is parallel to the beam (long) direction, $x$ points
in the direction of the total momentum of the pair in the locally co-moving system (LCMS) (out), and $y$ is perpendicular to the first two axes (side).

The correlation moments, for each order $l$, may be calculated from the measured 3D correlation function using Eq.~(\ref{eqn2}), but can be then vulnerable to directional inefficiencies.  Alternatively, the moments may be fitted using Eq.~(\ref{eqn1}), avoiding regions of poor efficiency or accounting for those in the fit weights.  In following the latter type of analysis, Eq.~(\ref{eqn1}) has been truncated at $l=4$ and expressed in terms of independent moments only.  The higher order moments have been found
to be negligible. Up to order 4, there are 6 independent moments:
$R^0$, $R^2_{x2}$, $R^2_{y2}$, $R^4_{x4}$, $R^4_{y4}$ and $R^4_{x2y2}$
where $R^2_{x2}$ is shorthand for $R^2_{xx}$ etc. These independent moments were extracted as a function of $q$, by fitting the truncated series to
the measured 3D correlation function with the moments as the parameters of
the fit.

\subsection{Source Reconstruction}

 Each independent correlation moment can be imaged using the 1D source imaging
code of Brown and Danielewicz \cite{brown97,brown98,brown01} to obtain the corresponding source moment within each order~$l$. Both the effects of Bose-Einstein symmetrisation and of Coulomb interaction, the sources of the observed correlations, are accounted for in the source imaging code.
Thereafter, the total source function, $S(\mathbf{r})$, can be reconstructed by combining the source
moments for each~$l$:
\begin{equation}
 S(\mathbf{r}) = \sum_l \sum_{\alpha_1 \ldots \alpha_l}
   S^l_{\alpha_1 \ldots \alpha_l}(r) \,A^l_{\alpha_1 \ldots \alpha_l} (\Omega_\mathbf{r})
\label{eqn3}
\end{equation}

Another method of construction of 
the 3D source function from the 3D correlation function is by fitting the latter directly upon assuming the shape for the source function. Since the 3D correlation function is represented by the Cartesian moments in
the Cartesian harmonic decomposition, this amounts to fitting the six independent moments, dependent on~$q$, with a trial source function. Two trial functional shapes were considered in this analysis: the 3D Gaussian, frequently termed an ellipsoid shape,
 and the Hump function, with six adjustable parameters, given by
\begin{equation}
  S(x,y,z) = \lambda \exp[-f_s(\frac{r^2}{4 r_{s}^2}) - f_l(\frac{x^2}{4 r_{xl}^2}
  + \frac{y^2}{4 r_{yl}^2} + \frac{z^2}{4 r_{zl}^2})]
  \label{ss_eqn}
\end{equation}
where $f_s=1/[1+(r/r_0)^2)]$ and $f_l = 1-f_s$.

\section{Results}

\begin{figure}
\setcaptionmargin{5mm}
\onelinecaptionstrue
\resizebox{21pc}{!}{\includegraphics{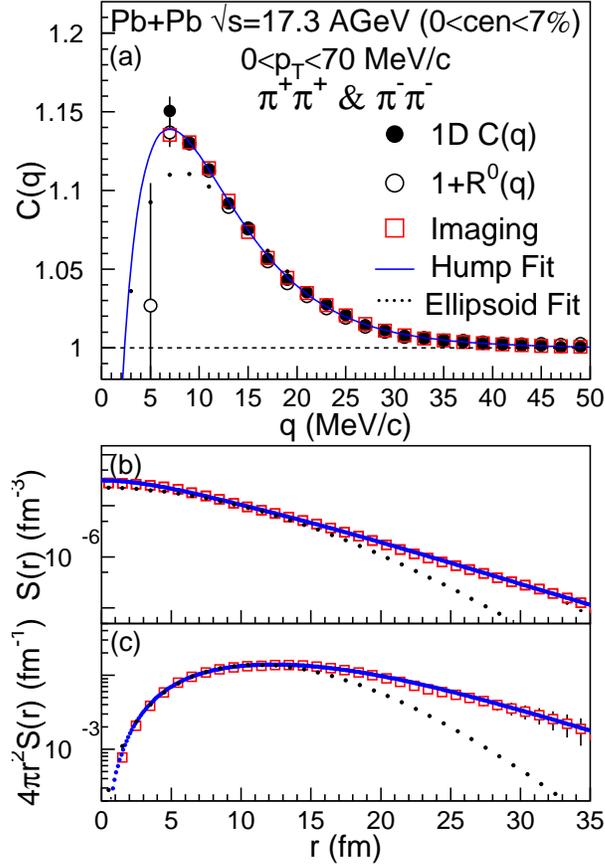}}
\vskip -0.5cm
\caption{\label{na49_fig1_ppg}
(color on line) Angle averaged correlation function (top panel) and source function
(middle) and radial probability density (bottom). Filled circles
show correlations from a direct angle-averaging of the data.  Open circles represent correlations
from fitting the angular decomposition to the data. Squares show the imaged source and
correlations corresponding to the imaged source. The dotted and solid lines represent,
respectively, the fitted Gaussian and Hump sources and their corresponding correlation
functions (see text).}
\end{figure}

Figure \ref{na49_fig1_ppg} (a) shows a comparison between the 1D correlation function C($q$) and the $l=0$ moment $1+R(q)$ for mid-rapidity low-$p_T$ pion-pairs from central Pb+Pb collisions. The~functions are in very good agreement with each other as expected in the absence of significant detection inefficiencies with dependence on angle in  $\mathbf q$-space. The figure also shows that the Hump function (solid line) fits the data very well while the ellipsoid fit (dotted line) underestimates the data at
 $q<13$~MeV/c. This difference in correlation fits at low $q$ corresponds to a difference in the source function at large $r \gtrsim 15$~fm, as evident in Fig.~\ref{na49_fig1_ppg} (b). The Hump shape (solid line) is in good agreement with the source image
 (squares) whereas the ellipsoid shape (dotted line) underestimates the image.
Given that the discrepancy occurs at large~$r$, that discrepancy becomes even more pronounced for the radial density in Fig.~\ref{na49_fig1_ppg}(c).

\begin{figure}
\setcaptionmargin{5mm}
\onelinecaptionstrue
\resizebox{30pc}{!}{\includegraphics{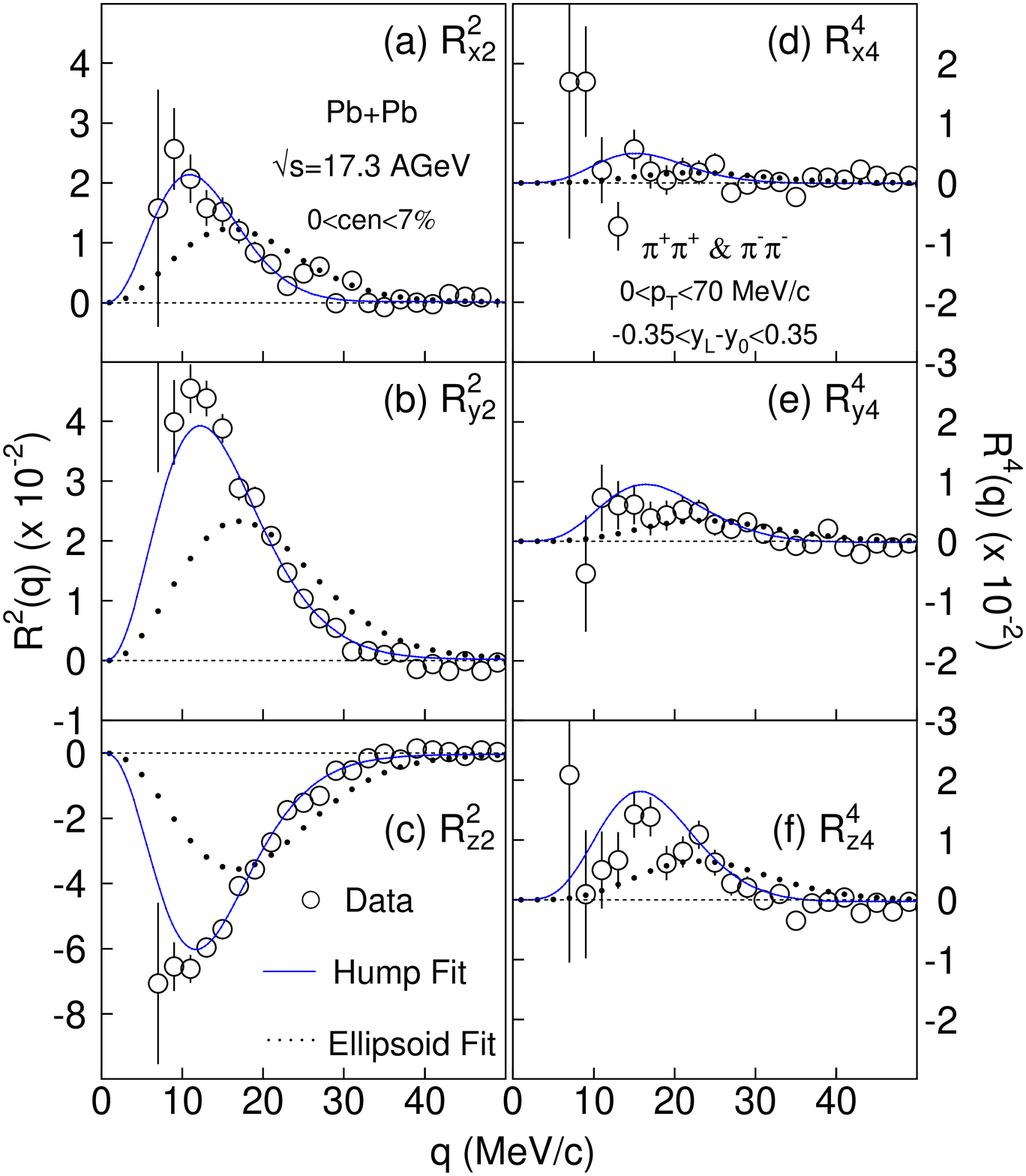}}
\vskip -0.75cm
 \caption{\label{na49_fig2_ppg}
{(color online) Correlation moments of multi-polarity $l=2$ (left panels),
and $l=4$ (right panels) for $\pi^+\pi^+$ and $\pi^-\pi^-$ pairs. 
Results are represented in the same way as in Fig.~\protect\ref{na49_fig1_ppg}, with error bars indicating statistical errors for the data.
Systematic errors are smaller than the data points.}
}
\end{figure}

Figure~\ref{na49_fig2_ppg} shows the different $l=2$ and $l=4$ moments (open circles) as a function of pion separation. The $l=4$ moments are significantly smaller in magnitude compared to the $l=2$ moments: this justifies the truncation of the series Eq.
~(\ref{eqn1}) at $l=4$. The Hump fit (solid line), with $\chi^2/\text{ndf} = 1.3$, is in close agreement with the data whereas the ellipsoid function (dotted line) gives a poor fit to the data ($\chi^2/\text{ndf} = 6.8$), as is visually evident.

\begin{figure}
\setcaptionmargin{5mm}
\onelinecaptionstrue
\resizebox{30pc}{!}{\includegraphics{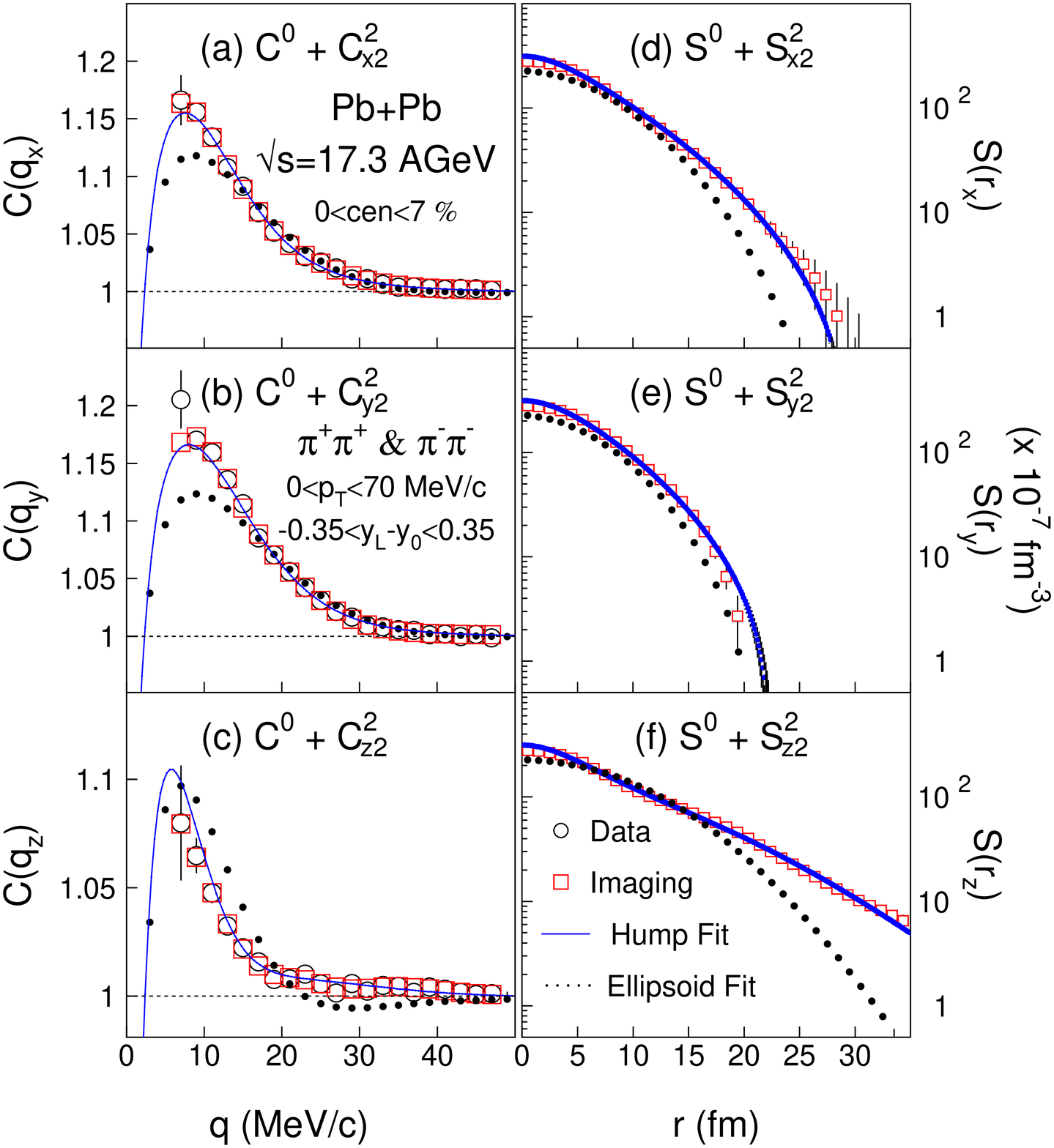}}
\vskip -0.75cm
 \caption{\label{na49_fig3_ppg}
{(color online) Correlation $C(q)$ (left panels) and source $S(r)$ (right panels) function profiles for $\pi^+\pi^+$ and $\pi^-\pi^-$ pairs in the outward $x$ (top panels), sideward $y$
(middle) and longitudinal $z$ (bottom) directions. The use of symbols is analogous to that
in Fig.~\ref{na49_fig1_ppg}. Error bars indicate statistical errors. Systematic
errors do not exceed the symbol size. Here, $l=4$ moments make negligible contributions.}}
\end{figure}

Figure~\ref{na49_fig3_ppg} (a)-(c) shows the 3D correlation function profiles in the $x$, $y$ and $z$ directions while the corresponding source function profiles are shown in panels (d)-(f). The source image (squares) is in good agreement with the Hump function fit (solid line), but is underestimated by the ellipsoid fit (dotted line) in all 3 directions. These non-Gaussian tails in the Hump source function profiles have corresponding manifestations in the correlation function profiles at low~$q$, relative to the ellipsoid fit.

\section{Therminator Model}

\subsection{Model Assumptions}

The Therminator model \cite{kis05} can shed more light
on the source breakup and emission dynamics. The Therminator model incorporates the following: (1) the Bjorken assumption of longitudinal boost invariance, (2) blast-wave expansion in the transverse direction with transverse velocity profile semilinear in
 transverse radius $\rho$ \cite{kis07}, (3) thermal emission of particles from a cylinder of infinite longitudinal size and finite transverse radius $\rho_{max}$, and (4) all known resonance decays.

 Assumptions (1) and (2) of the model imply that the expanding source
 consists of fluid elements shaped like onion rings with their
 axes aligned with the beam axis. Each fluid element expands
 transversely with a transverse velocity semi-linear in $\rho$,
 as well as translates longitudinally with a velocity profile
 linear in $z$.

 Each fluid element breaks up after a proper breakup time $\tau$
 in its own rest frame, with subsequent particle emission. Bjorken
 assumption of longitudinal boost invariance implies that the
 breakup (emission) time, $t$, of a source element at position $z$
 in the laboratory frame is given by $t^2 = \tau^2 + z^2$ .

 At the point of source breakup, particles within the Therminator model leave 
 each source element defined by their longitudinal and
 transverse positions: $z$ and $\rho$. All particle emissions are
externally viewed as occuring from a freeze-out
 hypersurface defined in $\rho$-$\tau$ plane as $\tau = \tau_0 + a \rho$,
 where $a$ is a parameter which controls the space-time correlation
 and $\tau_0$ is the proper breakup time of the source element
 located at $\rho=0$, i.e.\ on the beam axis,
 and $\tau$ is proper breakup time of the source element located
 at finite $\rho$. Hence, particles which are emitted from a generic fluid
 element with coordinates ($z$,$\rho$) will have emission time $t$ in the laboratory frame
 given by $t^2 = (\tau_0 + a \rho)^2 + z^2$ .

 In this analysis, Therminator is used in the blast-wave mode and $a$ is set to the negative value of 
$-0.5$ \cite{kis06}. Hence, $\tau$ decreases linearly with
 $\rho$ down to a minimum value at $\rho=\rho_{max}$. The negative
 value of $a$ implies a negative space-time correlation, i.e.\ particles
 at large $\rho$ freeze-out earlier. Hence, the source emits or
 burns from outside in.

 Since each source element is defined by only one value of the
 proper breakup time $\tau$, all particle emissions from this
 fluid element happen instantaneously in the rest frame of the
 source element and the proper emission duration in the above scenario is 0.
 On the other hand, one can allow for a finite non-zero proper
 emission duration in each source element, if the single proper
 breakup time $\tau$ of each source element is replaced with a
 distribution of breakup times. This corresponds physically to
 the source element breaking up over a finite time interval
 rather than instantaneously.

 One such parametrization is an exponential
 distribution of breakup times $\tau'$ with a width $\Delta\tau$ given by
 $dN/d\tau' = \frac{\Theta(\tau'-\tau)}{\Delta\tau} \, \text{exp}[-(\tau' - \tau)/\Delta\tau]$ . The minimum value of $\tau'$ is the initial single breakup time $\tau$
for that source element. With such a distribution of breakup times, each
 source element emits particles from a family of hypersurfaces,
 each of which is defined by a $\tau'$ value which is sampled
 according to the exponential distribution. In such a scenario,
 particle emission from each source element takes place over
 a finite time duration defined by $\Delta\tau$ rather than
 instantaneously. In this parametrization, $\Delta\tau$ represents the proper emission duration in the rest frame of the fluid element.

\subsection{Comparison to Data}

\begin{figure}
\setcaptionmargin{5mm}
\onelinecaptionstrue
\resizebox{25pc}{!}{\includegraphics{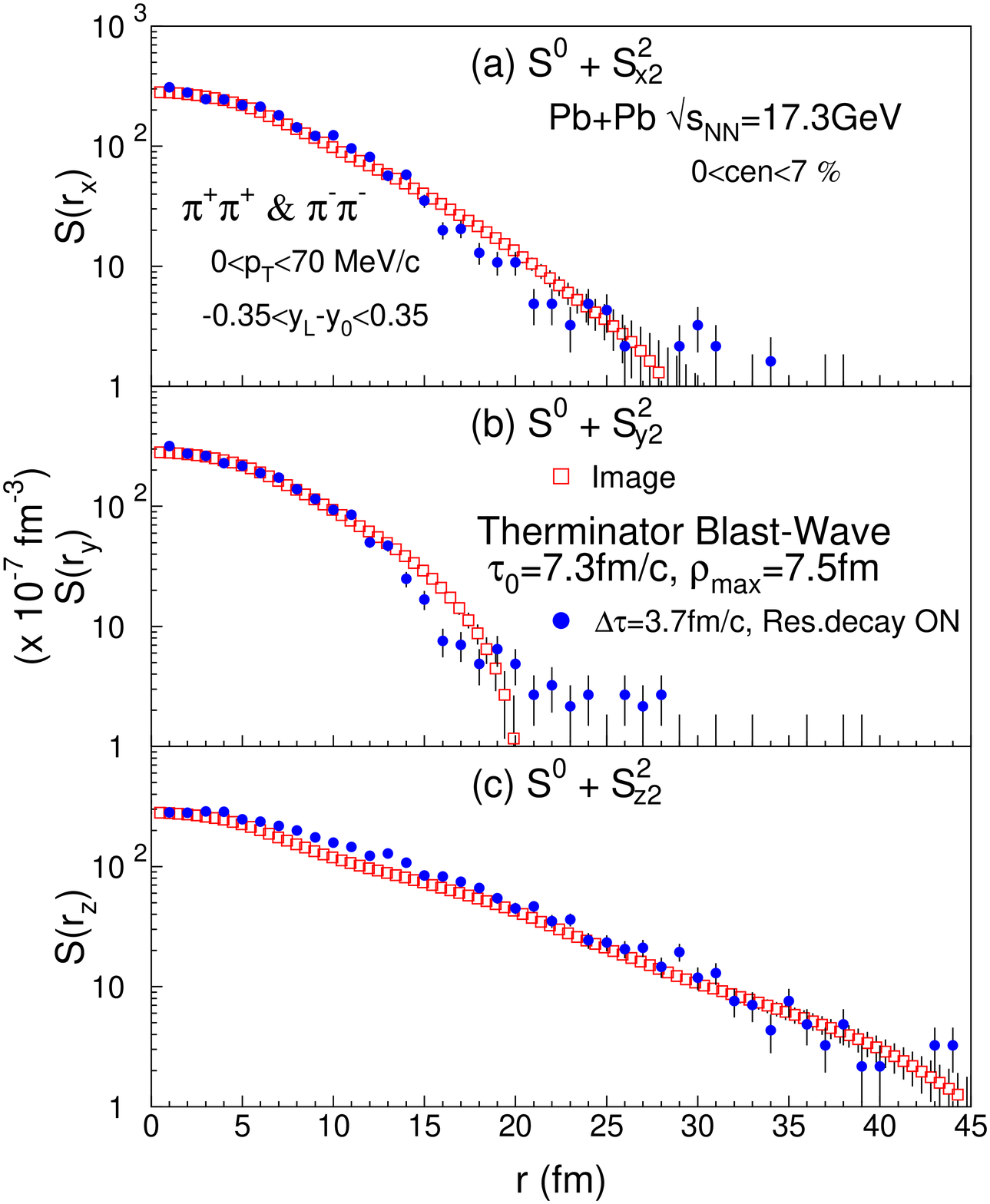}}
\vskip -0.75cm
 \caption{\label{na49_fig4_ppg}
{Source function profiles, $S(r_i)$, along $x$ (a), $y$ (b) and $z$ (c) directions, from imaged correlation function (squares) and from the Therminator model (circles).}
}
\end{figure}

Figure~\ref{na49_fig4_ppg} shows a comparison of the source image (squares) with the calculated source function (circles) from the Therminator model in its blast-wave mode. A good match is obtained with values of $\tau_0=7.3$~fm/c and $\Delta\tau=3.7$~fm/
c, when all resonance decays are turned on. The good agreement between extracted source image and calculated source function indicates that the data are consistent with the basic ingredients of Bjorken longitudinal expansion and with the blast-wave dynamics for transverse expansion which are incorporated in the Therminator model. It also indicates that a finite non-zero pion emission duration and outside-in burning of the source are needed to describe the data.

\section{Conclusions}

The model-independent 3D source imaging technique, involving decomposition of the 3D correlation function into Cartesian surface-spherical harmonics, has been applied to pion pairs from Pb+Pb collisions at $\sqrt s_{NN}=17.3$~GeV. Prominent non-Gaussian tails have been observed in the outward direction of the pion pair transverse momentum and in the longitudinal direction of the beam. The extracted source image is well described by the Therminator model which incorporates Bjorken longitudinal expansion and blast-wave transverse-flow dynamics. The data is consistent with a proper emission duration of 3.7~fm/c for pions at $\sqrt s_{NN}=17.3$~GeV.

\end{document}